\documentclass[two column,prl]{revtex4}
\usepackage{bm}
\usepackage{graphics}
\usepackage{graphicx}
\usepackage{amsfonts}
\usepackage{amsmath}
\usepackage{amssymb}
\usepackage{graphicx}

\begin{document}

\title[Short title for running header]{White Light Interferometry for Quantitative Surface Characterization in Ion Sputtering Experiments}

\author{S.V. Baryshev}
\email{sergey.v.baryshev@gmail.com} \affiliation{Materials Science
Division, Argonne National Laboratory, 9700 S. Cass Avenue,
Argonne, IL 60439, USA}

\author{A.V. Zinovev, C.E. Tripa}
\affiliation{Materials Science Division, Argonne National
Laboratory, 9700 S. Cass Avenue, Argonne, IL 60439, USA}

\author{R.A. Erck}
\affiliation{Energy Systems Division, Argonne National Laboratory,
9700 S. Cass Avenue, Argonne, IL 60439, USA}

\author{I.V. Veryovkin}
\affiliation{Materials Science Division, Argonne National
Laboratory, 9700 S. Cass Avenue, Argonne, IL 60439, USA}


\begin{abstract}
White light interferometry (WLI) can be used to obtain surface
morphology information on dimensional scale of millimeters with
lateral resolution as good as $\sim$1 $\mu$m and depth resolution
down to 1 nm. By performing true three-dimensional imaging of
sample surfaces, the WLI technique enables accurate quantitative
characterization of the geometry of surface features and compares
favorably to scanning electron and atomic force microscopies by
avoiding some of their drawbacks.

In this paper, results of using the WLI imaging technique to
characterize the products of ion sputtering experiments are
reported. With a few figures, several example applications of the
WLI method are illustrated when used for (i) sputtering yield
measurements and time-to-depth conversion, (ii) optimizing ion
beam current density profiles, the shapes of sputtered craters,
and multiple ion beam superposition and (iii) quantitative
characterization of surfaces processed with ions.

In particular, for sputter depth profiling experiments of
$^{25}$Mg, $^{44}$Ca and $^{53}$Cr ion implants in Si
(implantation energy of 1 keV per nucleon), the depth calibration
of the measured depth profile curves determined by the WLI method
appeared to be self-consistent with TRIM simulations for such
projectile-matrix systems. In addition, high depth resolution of
the WLI method is demonstrated for a case of a Genesis solar wind
Si collector surface processed by gas cluster ion beam: a 12.5 nm
layer was removed from the processed surface, while the transition
length between the processed and untreated areas was 150 $\mu$m.

\end{abstract}

\maketitle

\section{Introduction}
\label{intro}

In many experiments designed to determine sputtering yields (SY)
of various materials under specific ion bombardment conditions,
uncertainties in ion beam parameters can propagate and result in
uncertain sputtering yield values \cite{1}. For example, it can be
challenging to determine shapes of ion beam profiles and the
corresponding operational current densities, especially when the
projectile energy goes below 1 keV and then further approaches the
sputtering threshold. Moreover, under such conditions, the
focusing of the ion beam is in question, and the relative spread
$\Delta\varepsilon/\varepsilon$ in the initial kinetic energy
distribution of ions \cite{2} can have strong influence on
experimental results \cite{3,4}.

The other aspect that has a great impact on the final results is
the method used for quantitative analysis of the surface, being
commonly scanning electron and atomic force microscopy (SEM and
AFM, respectively). Both techniques are valuable, but each has its
own limitations, when used for surface morphology
characterization. The AFM can obtain three-dimensional (3D)
imaging and thus the cross section profiles for sputtering
craters, but AFM has rather narrow ranges in the maximum lateral
and especially depth scanning. The SEM has much greater
flexibility in the size of field-of-view with large depth of
focus, but obtaining 3D imaging is cumbersome \cite{5}. Another
technique widely used in secondary ion mass spectrometry is the
Stylus Profilometry. This technique is popular because of its
simplicity, but it is a coarser contact tool able to scan along a
single line at a time, which would make 3D surface imaging
extremely time consuming. The qualifier "coarser" means the Stylus
has difficulty measuring surface features of high aspect ratio or
of size comparable with its characteristic tip size that implies a
tip radius along with a tip angle \cite{6}. It should be mentioned
that in the case of the trace analysis mass spectrometry (our
case), it is undesirable to have a sample to be analyzed in
physical contact with a Stylus tip, which may contaminate or even
scratch the surface. All these facts make researchers to look for
alternative methods for surface topography measurements. In this
regard, the optical interference methods seem to be natural. It is
known that the main drawback of an optical technique (utilizing
geometrical optics) is the limited lateral resolution against SEM
and AFM. This limitation is of fundamental nature in that a
surface feature of characteristic size less than $\sim\lambda/2$
(where $\lambda$ is a light wavelength) cannot be resolved
correctly. On the other hand, the interference approach gives a
fascinating depth resolution of less than 1 nm.

This work reports on application of the white light profilometry
based on a Mirau interferometer (which is common for most of the
commercial instruments) to characterize solid surfaces eroded in
ion sputtering experiments. A few examples of applying this method
are provided when used for (i) characterization of ion beam
profiles and crater shapes yielding accurate SY estimates, (ii)
overlap alignment of a multiple ion beams system, (iii)
time-to-depth calibration in sputter depth profiling, and (iv)
characterization of surface processing of materials by ion beams.
For sputtering yield and rate estimates, the presented results
demonstrate an alternative experimental approach to generate
reference data for many materials and technological applications
\cite{4,7,8,9,10,11,12,13} under bombardment with both commonly
used atomic ions and relatively new molecular and cluster ions and
help to resolve the problem of time or primary ion fluence to
depth conversion.

In regard to mass spectrometry experiments, investigation of WLI
benefits is practically important for us, since WLI as non-contact
optical technique is attractive for implementing as an
\emph{in-situ} characterization tool.

\section{Material and methods}
\label{sec:2}

Mirau interferometry is an optical technique that measures the
phase shift between the reference light signal and the light
reflected from the sample surface. It provides an optical
micrograph onto which constructive and destructive interference
fringes (light/dark) are superimposed. The fringes are used to
reconstruct the three-dimensional surface profile. A white light
source supplies a broad spectrum light. This eliminates the
problem associated with certain specimen features where the
correct interference order cannot be determined. The lateral
resolution of the WLI probe is determined mostly by the chosen
numerical aperture of the objective (limited to $\sim\lambda/2$ at
numerical aperture $\sim$1). Once the best focus is found by
mechanical positioning of the sample stage and the objective
(corresponding to the brightest and strongest interference
fringes, see Fig.1), a piezo transducer inside the objective
performs vertical scanning of heights over a specified range. Then
an array of phase shifts between the reference signal, with
constant optical path, and the signal, with an optical path which
depends on the depth, is used to reconstruct true 3D surface
topography and morphology. At first glance, it seems that
optically transparent films on a reflective substrate pose a
serious problem for WLI. If a material is transparent for given
wavelength $\lambda$, there is always a phase shift (optical path
length change) due to multiple passes of the light inside a film
of refractive index $n>1$, which may yield artifacts in a 3D
topographic image. At the same time, the phase shift allows one to
distinguish between the transparent film response and the signal
originated from the reflective base. By separating these two
responses (either directly \cite{14} or by a special
post-processing algorithm \cite{15}) and paying attention to an
absorption characteristic (which can be obtained independently)
\cite{16}, one can measure a transparent/semitransparent film
thickness starting at an order of 10 nm or higher (up to several
$\mu$m), so that the drawback may turn out to be an additional
advantage. Information on Mirau WLI can be found in
Refs.\cite{17,18} in great details.

In the experiments presented here a MicroXAM-1200 profilometer
controlled via MapVue AE software was employed. The images were
visualized using the SPIP software. Before every measurement, the
profilometer was calibrated laterally by a precise sub-mm ruler
and vertically by 500 nm step AFM standard from Ted Pella, Inc.

In the examples of application of the WLI to ion sputtering
experiments that follow, small ($\sim$10$\times$10 mm$^2$) pieces
of Si(001) (MEMC Electronics and Unisil), and Cu(110) and Cu(111)
(MTI Corporation) monocrystals were utilized. In this context,
these Si and Cu samples (which are uniform and nontransparent
materials) do not have the "transparent sample" problem described
above. In addition, it seems that in many sputtering experiments,
including the present study, shapes of removed craters, spots,
etc. can be classified as low gradient or step-like, which favors
WLI applicability \cite{19}.

\begin{figure}[t] \centering
\includegraphics[width=5.5cm]{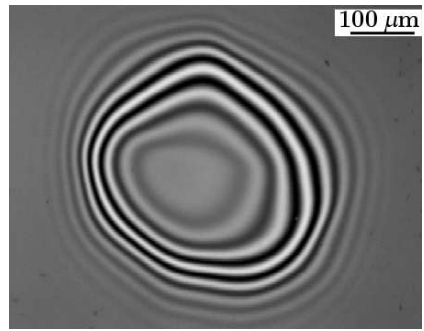}
\caption{Optical micrograph showing example of the
optimally-aligned interference fringes from a white light
profilometer. The sample is a sputtering semispherical crater
formed by direct current ion irradiation of a small Si wafer
chip.}\label{fig:1}
\end{figure}

\section{Results and discussion}
\label{sec:3}

\

\emph{1. Craters and ion beam profiles measurements to estimate
Sputtering Yield}

As an alternative to the known and widely employed method for
estimating sputtering yields using mass-loss method, based on
direct weighing or quartz microcrystal balance \cite{20}, we
propose to use the WLI method for direct visualization of the
sputtered ion beam spots or craters obtained by static sputtering
or by raster scanning of an ion beam, respectively. For low energy
ion beam irradiation, WLI can verify whether or not the entire
beam was confined to the sample of interest. By combining the WLI
visualization with precise measurements of the total ion current
by a Faraday cup, the SY and the operating current density can be
obtained simultaneously. Besides, this approach appears to be very
helpful in estimating the extent of undesirable "wings" of the ion
beam profile so that, as a feedback, it guides the alignment of an
ion beam source. The sputtering yield $Y$ is then estimated using
the following expression

\begin{equation}
Y=\frac{\rho\cdot V\cdot e}{I\cdot\tau\cdot M_{atom}}, \label{1}
\end{equation}
where $I$, direct current (dc) current of an ion beam; $\tau$,
time of sputtering; $M_{atom}$, mass of a matrix atom in grams;
$\rho$, density; $e$, the elementary charge. $V$ is the volume of
the removed sample material obtained by means of the WLI
measurement. Volume calculations can be performed either by using
a histogram of heights typically available through an
interferometer post-processing software called SPIP by Image
Metrology that works with files type generated by MapVue AE or by
three dimensional integration based on cross sections in two
orthogonal directions centered on the eroded surface area (black
lines in Fig.2a).

Figure 2 compares longitudinal cross sections of a spot (red
dotted line) of a normally incident static 5 keV Ar$^+$ ion beam
against a crater (green open squares) obtained by 100$\times$100
pixels digital raster scanning of the same ion beam over the
surface of a Cu(110) monocrystal. The curve corresponding to the
static beam overlaps one edge of the crater to demonstrate how
raster scan of the ion beam generates the crater during sputter
depth profiling. Good alignment of the ion beam column manifests
itself in a symmetric beam profile and FWHM of 120 $\mu$m at a
total current of 2 $\mu$A. The WLI approach allows one to
characterize the ion sputtering with the same normally incident
ion beam decelerated to 150 eV by the target potential. In this
case, the cross section of the static beam spot is shown by an
orange solid line, and the crater cross section is shown by cyan
open circles. The ion column allowed delivery of the same 2 $\mu$A
of Ar$^+$ current on the target because the deceleration of the
beam from the nominal 5 keV energy to 150 eV occurred in the
immediate vicinity of the target, and in such a way that its
optimal focusing was maintained by an electrostatic lens (FWHM of
150 $\mu$m in Fig.2b proves that) \cite{21}. The sputtered crater
has in this case a larger lateral size because the deflection
voltages of the raster-generating octupole were kept unchanged for
the two primary ion impact energies, resulting in additional beam
swinging due to the target potential.

Based on the WLI data, sputtering yields of Cu(110) at 5 keV and
150 eV ion impact energies were determined. An obtained SY value
of 1.8 at/ion for the former case was in good agreement with
literature data \cite{22}. For the latter one, the sputtering
yield was 0.2 at/ion. The SY values for Cu(111) at 50, 100, and
150 eV were also determined as 0.13, 0.27, and 0.42 at/ion,
respectively. The measured energy spread $\Delta\varepsilon$ of
the low energy system \cite{21} is 23 eV.

\begin{figure}[t] \centering
\includegraphics[width=5cm]{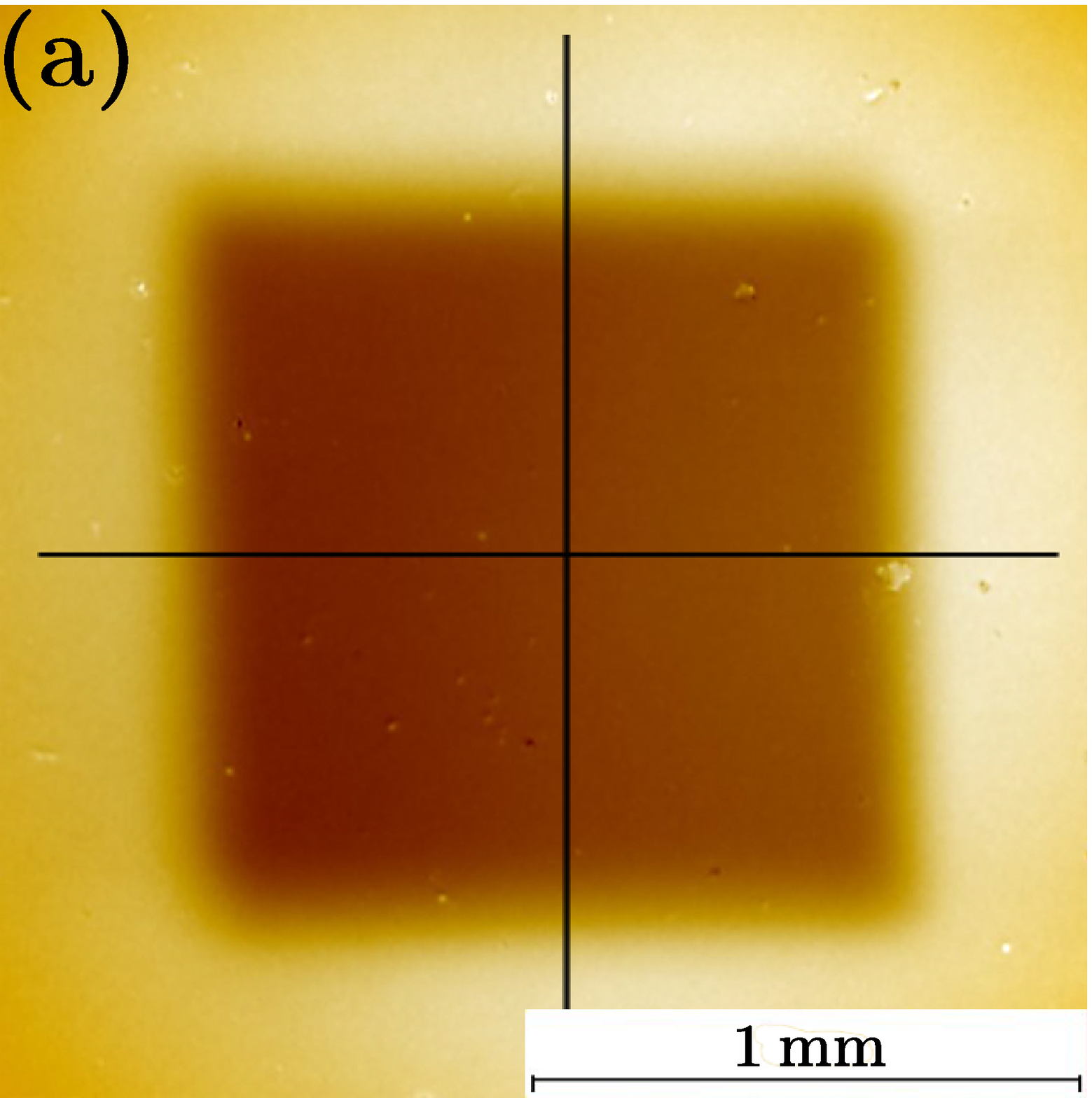}

\

\includegraphics[width=6cm]{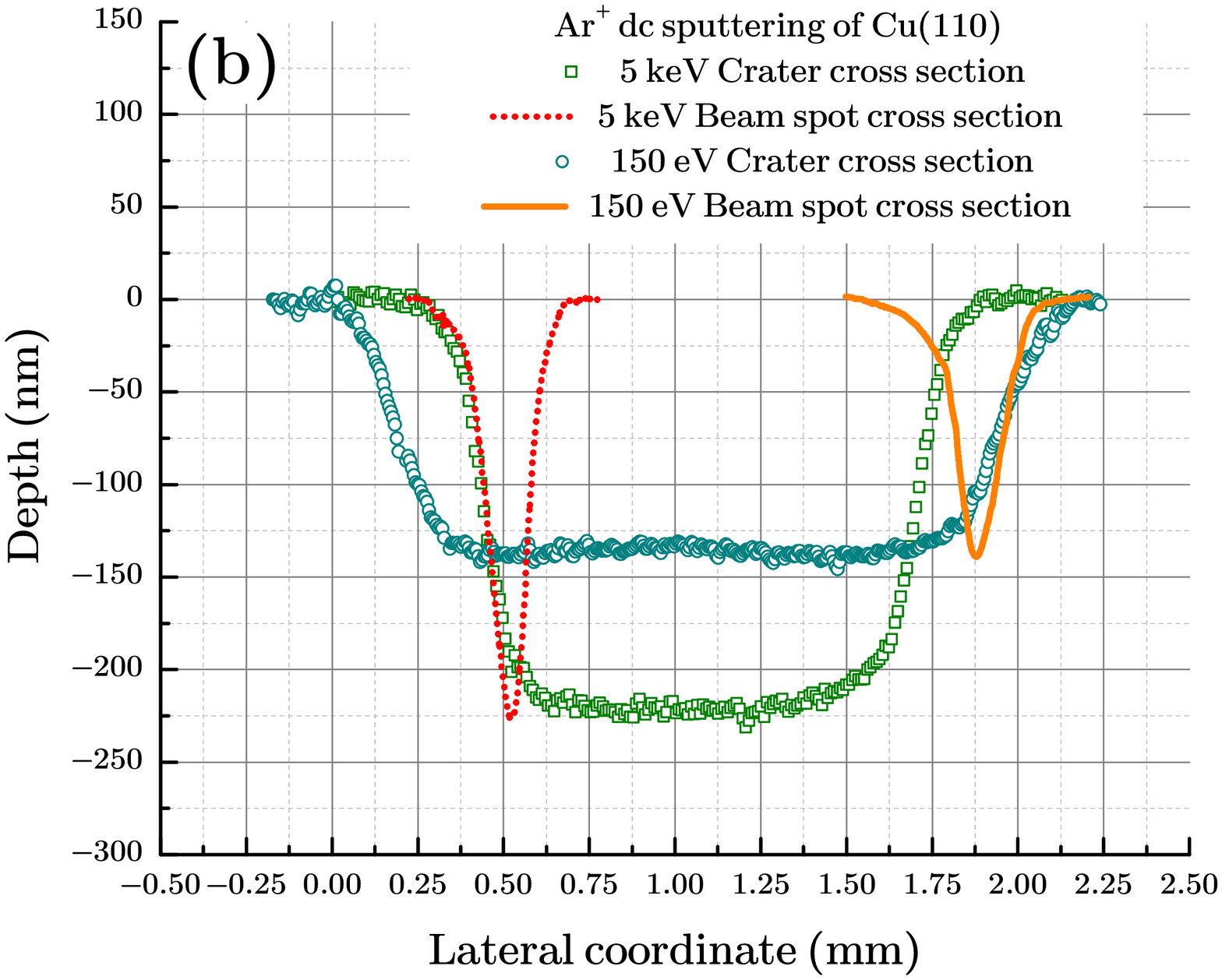}
\caption{(a) Pseudocolor 2D top view of produced crater. Black
lines are directions along which cross sections plotted in (b)
were measured. (b) Beam spot and crater cross sections
superimposed. Measurements were made on Cu(110) sputtered by
normally incident Ar$^+$ ion beams with 5 keV (green squares and
red dotted line) and 150 eV (cyan circles and orange solid line)
energies.} \label{fig:2}
\end{figure}

\

\emph{2. Multiple beam system alignment and time-to-depth
conversion}

In our previous work, we have introduced and demonstrated a new
variant of dual-beam (DB) sputter depth profiling for
time-of-flight secondary ion mass spectrometry (TOF SIMS), where
we aimed at improving the depth resolution by using a normally
incident low-energy direct current ion beam for sputtering, in
combination with obliquely incident fine focused pulsed ion beam
for TOF SIMS analysis. The benefit of such an arrangement of the
sputtering ion beam is two-fold: its low (a few hundred eV) energy
reduces ion beam mixing, and its normal impact angle reduces
surface roughening. To make this concept work, it is needed to
precisely overlap the crater created by raster scanning the low
energy dc ion beam with the area probed by the pulsed analysis ion
beam. Moreover, (i) most of the bottom of the low energy crater
must be flat (Fig.2b), and (ii) the analysis area must be confined
within that flat part, in order to avoid distortions in the depth
profile due to probing sloped areas or crater walls. This can only
be achieved by thorough optimization of both ion beams (current
density profiles and focusing) as well as precise control of their
steering. The WLI technique helps to make this multi-step
alignment much easier.

Results of the WLI characterization presented in Fig.3 give
straightforward answers regarding mutual positioning of sputtering
and analysis ion beams by showing two craters produced by raster
scanning of these beams in dc mode. The deep and narrow crater
seen in Fig.3 was made by the analysis beam (5 keV Ar$^+$ ions
with 60$^\circ$ incident angle). The wide and shallow crater was
made by a normally incident 500 eV Ar$^+$ ion beam. Fig.3
demonstrates that the 5 keV Ar$^+$ probing in the DB mode was
conducted on the flat bottom part of the crater created by the low
energy sputtering ion beam.

\begin{figure}[t] \centering
\includegraphics[width=5cm]{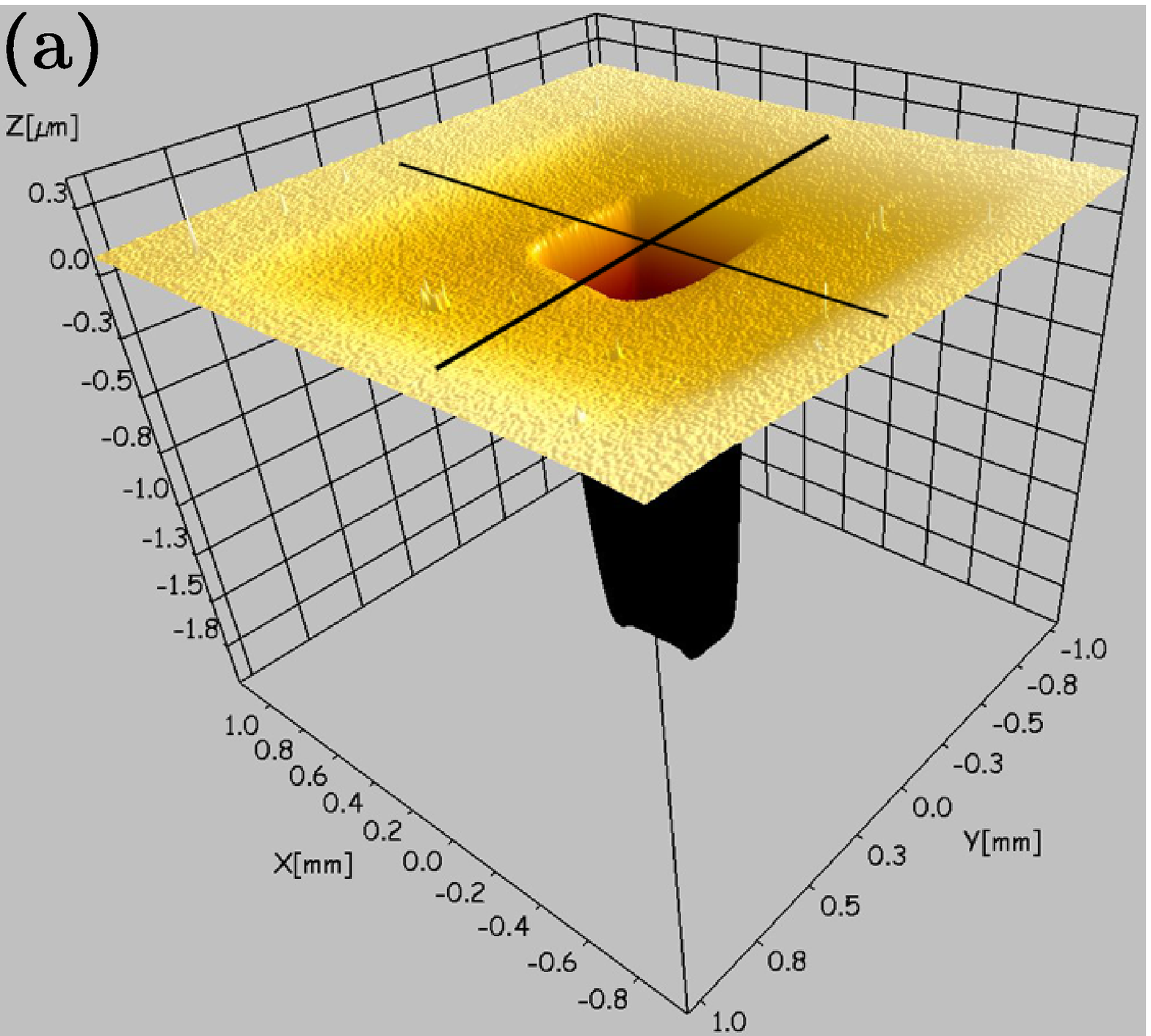}

\

\includegraphics[width=6cm]{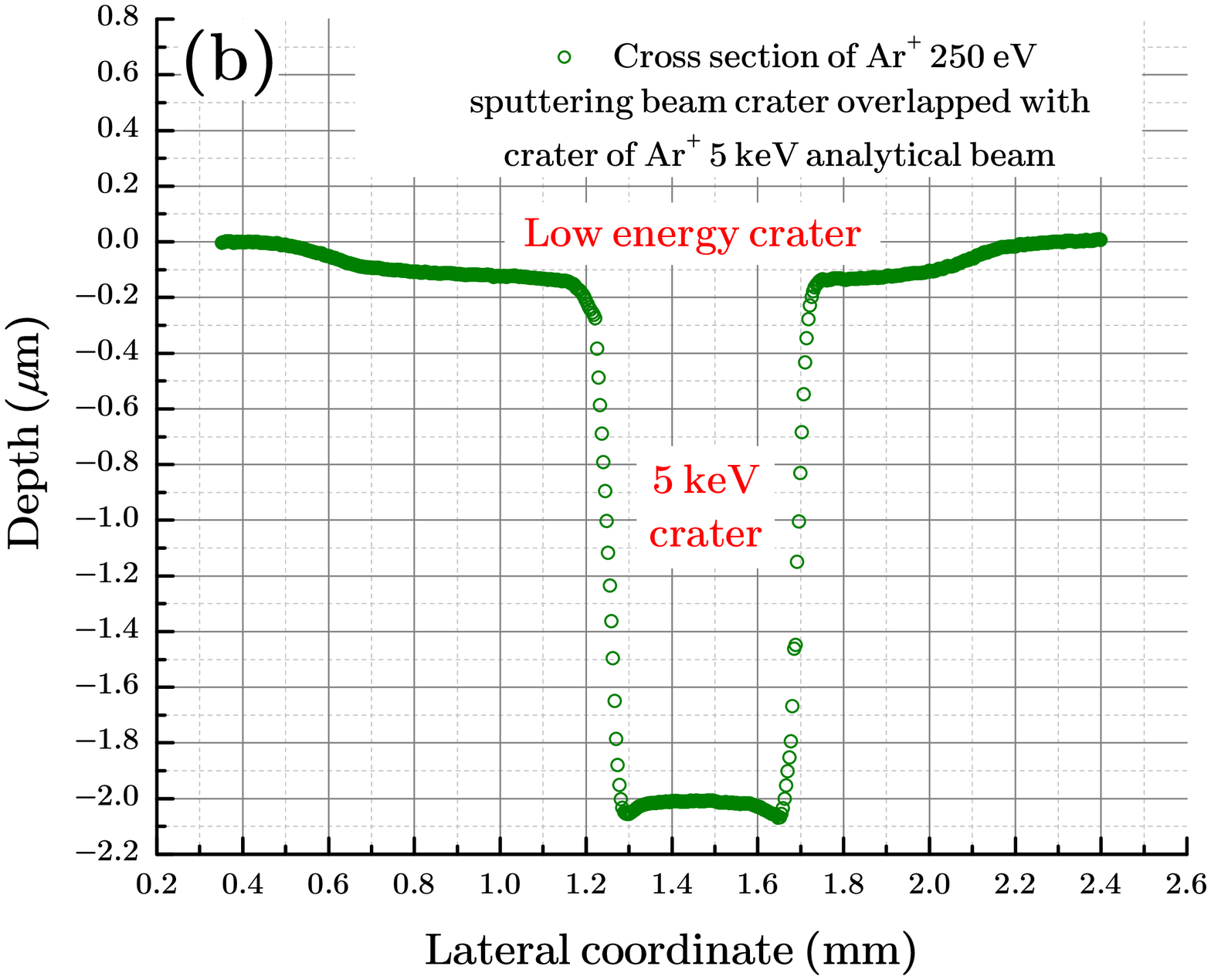}
\caption{(a) Pseudocolor 3D topographic view of two superimposed
craters made by two separate raster scanned ion beams. The large
one (1.5$\times$1.5 mm$^2$) is sputtered by a 500 eV normally
incident Ar$^+$ ion beam. The smaller one (500$\times$500
$\mu$m$^2$) is produced by 5 keV Ar$^+$ ions with 60$^\circ$
incidence angle. (b) Cross section of 3D image along one of the
black lines shown in (a)}
\label{fig:3}
\end{figure}

Another important application of the WLI method to sputter depth
profiling is exemplified by the sputtering time to sputtered depth
calibration procedure applied to this particular experiment. The
samples analyzed here were pieces of Si(001) wafer implanted with
$^{25}$Mg$^+$, $^{44}$Ca$^+$ and $^{53}$Cr$^+$ ions at energy of 1
keV per atomic mass unit (25 keV for $^{25}$Mg, 44 keV for
$^{44}$Ca and 53 keV for $^{53}$Cr, all at 3$\times$10$^{13}$
ions/cm$^2$ fluence) fabricated by Leonard Kroko Inc. A TOF MS
analysis of these samples was performed by laser post-ionization
of sputtered neutrals (secondary neutral mass spectrometry, SNMS)
using resonantly enhanced multi-photon ionization to
simultaneously detect all isotopes of Mg, Ca and Cr \cite{23}.
This was an experiment on sputter depth profiling which started in
the DB mode as described above but, after the concentration peaks
of the implants were passed (that is, after 170 nm on the depth
scale in Fig.3, see also Fig.4), the experiment continued in the
single beam (SB) mode by switching off the low energy sputtering
beam, while the analysis beam performed both the ion milling (in
dc mode) and the analysis (in pulsed mode). The higher energy (5
keV) and 60$^\circ$ incidence of the analysis beam allowed us to
reduce the time needed for measuring the trailing edge of the
implant depth profiles where high depth resolution was not needed.
The calibration procedure involved: (i) the WLI measurements of
the depths of craters created by both ion beams, as shown in
Fig.3, (ii) ion current measurements of both these beams with the
Faraday cup, and (iii) calculating depth scale based on the total
sputtering time with either of the two beams and the corresponding
WLI measurements of crater depths. To compare this depth
calibration with a model estimate, TRIM simulations for 1 keV/amu
ions of the same Mg, Ca and Cr isotopes implanted in a Si matrix
with SiO$_2$ of 2 nm on top were performed. After that, the
experimental and simulated data were compared on the same plot, as
shown in Fig.4. This comparison revealed very good agreement
between the depths of Ca and Mg implant peak concentrations
determined by the WLI-based depth calibration and simulated by
TRIM. In the case of Cr, the shift between simulated and
experimentally measured peak was $\sim$5 nm. Thus, the sputtering
time to depth calibration using the WLI measurements proved to be
satisfactorily accurate. It proved also that, if a depth profile
is made purely in SB manner, an elemental peak distribution
appears to lie deeper (under the same time-to-depth conversion
procedure by WLI) as compared to DB results shown in Fig.4. This
peak depth overestimation leads to an error in the fluence value
obtained by integration of the depth profile curve. This issue is
not discussed here, since this fact is obvious and lies beyond the
scope of this paper.

\begin{figure}[t] \centering
\includegraphics[width=6cm]{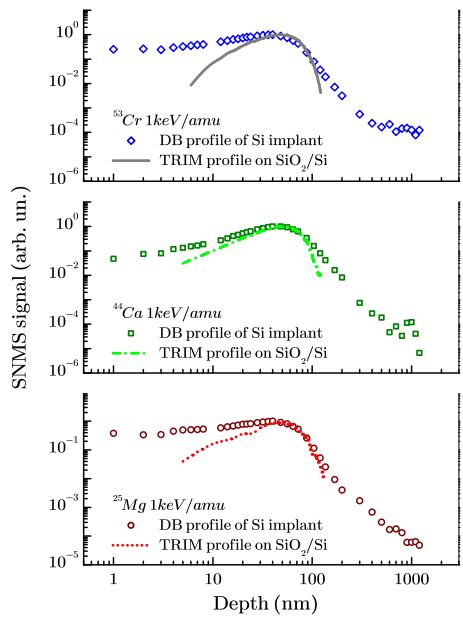}
\caption{Symbols represent measured secondary neutral
mass-spectrometry depth profiles of isotopes $^{25}$Mg (wine
circles), $^{44}$Ca (green squares), $^{53}$Cr (blue diamonds)
implanted in Si host matrix at 1 keV per nucleon (1 keV/amu).
Lines (red, light green and gray) are independent TRIM simulations
of depth distributions for the same isotopes of 1keV/amu energies
in SiO$_2$/Si sandwich (SiO$_2$ thickness is of 2 nm)}
\label{fig:4}
\end{figure}

\

\emph{3. Quantitative characterization of ultra-shallow surface
processing with cluster ion beam}

In this example, the depth resolution of the WLI technique applied
to characterization of Si surfaces irradiated with gas cluster ion
beams (GCIB) is demonstrated. The GCIB in these sputtering
experiments was an argon cluster beam Ar$^+_N$ with $N$=2000,
where $N$ corresponds to the number of atoms in the peak
distribution and, in general, can lie between 200 and 10000
\cite{24,25}. Irradiating materials surfaces with such cluster
ions causes two unique effects. First, because the impact energy
of such projectiles equals to their kinetic energy divided by the
number of constituent atoms, for a 20 keV Ar$^+_{2000}$, for
example, it will be only 10 eV/Ar, which significantly reduces the
penetration of individual Ar atoms into the target and the
sputtering process starts to strongly depend on the collective
effects of many such impacts. In essence, for GCIB irradiation,
the sample damage is confined to a narrow near-surface layer.
Another effect is the surface "polishing" (or planarization),
which manifests itself in a reduced roughness of the irradiated
surfaces. To summarize, at normal incidence the GCIB irradiation
can literally "shave off" topmost layers from a target with
minimal alteration of underlying regions.

These two effects are very beneficial for our efforts on
quantitative analyses of the Genesis mission \cite{26} solar wind
(SW) collectors by resonance ionization mass spectrometry
\cite{23}. The Genesis mission samples present a serious
analytical challenge because of abundant contamination which
blanketed the collectors surface after the crash landing of the
Genesis sample return capsule. In addition to the crash-derived
contamination, such as terrestrial dust particles, a highly
refractory organic/silicon film, known as the "brown stain"
\cite{27}, covers the top of Genesis samples. While conventional
methods such as megasonic cleaning with ultrapure water removes
particulates $\geq$1 $\mu$m loosely connected to surface
\cite{28}, the remaining contamination must be dealt with
differently. The GCIB processing of surfaces of Genesis collectors
has the potential to "shave off" this contamination blanket with
minimal losses of the implanted SW species \cite{29}. To our
knowledge, this is possibly the most advanced cleaning method
proposed so far for uniform removal of surface contamination.

In this WLI example the GCIB processed surface of the Genesis
60428 Si coupon is characterized in order to measure the exact
depth removed. Currently, by measuring $^{24}$Mg, $^{40}$Ca and
$^{52}$Cr solar wind distributions by DB SNMS, we know that the
surface contamination covers the first $\sim$10 nm of the depth
profile \cite{30}. By using the GCIB process to reduce the surface
contamination, the contribution of contamination to the depth
profile is significantly decreased, resolving the SW profile from
it and permitting a more accurate integration of the SW depth
profile curve to obtain elemental abundance fluences. Thus, the
precise thickness of the layer removed by GCIB is critical.

GCIB processing conditions on the Genesis 60428 Si coupon were as
follows: operating current of 68 $\mu$A, GCIB raster area of
6.4$\times$10$^{-3}$ m$^2$, GCIB exposed Si surface area of
2.9$\times$10$^{-5}$ m$^2$, and the sample processing time under
GCIB $T=\frac{2.9\times 10^{-5}}{6.4\times 10^{-3}}\times 153$ s
(where 153 s is the total time during which GCIB source was
switched on and raster scanned).

The measurement depicted in Fig.5 shows that the surface layer
removed by GCIB irradiation was as low as 12.5 nm. At the same
time, the length of the transition region between irradiated and
non-irradiated areas of the sample is as long as $\sim$150 $\mu$m.
This length is on the order of the full lateral scan of an AFM,
and makes it essentially impossible to find such a step by means
of AFM, while the depth is at the resolution limit of the best
Stylus Profilometer, emphasizing the high value of the WLI method.
If we assume that, originally, the sample consists of only Si and
use the literature data for sputtering yield of Si under a 20 keV
Ar$^+_{2000}$ cluster ion beam ($Y$=41.5 atoms per cluster ion
\cite{31}), the thickness that should have been removed would be
8.5 nm. This estimate proves indirectly the presence of an extra
layer that may contain submicron particulates, the "brown stain",
and the native silicon oxide layer before the GCIB processing.

\begin{figure}[t] \centering
\includegraphics[width=5cm]{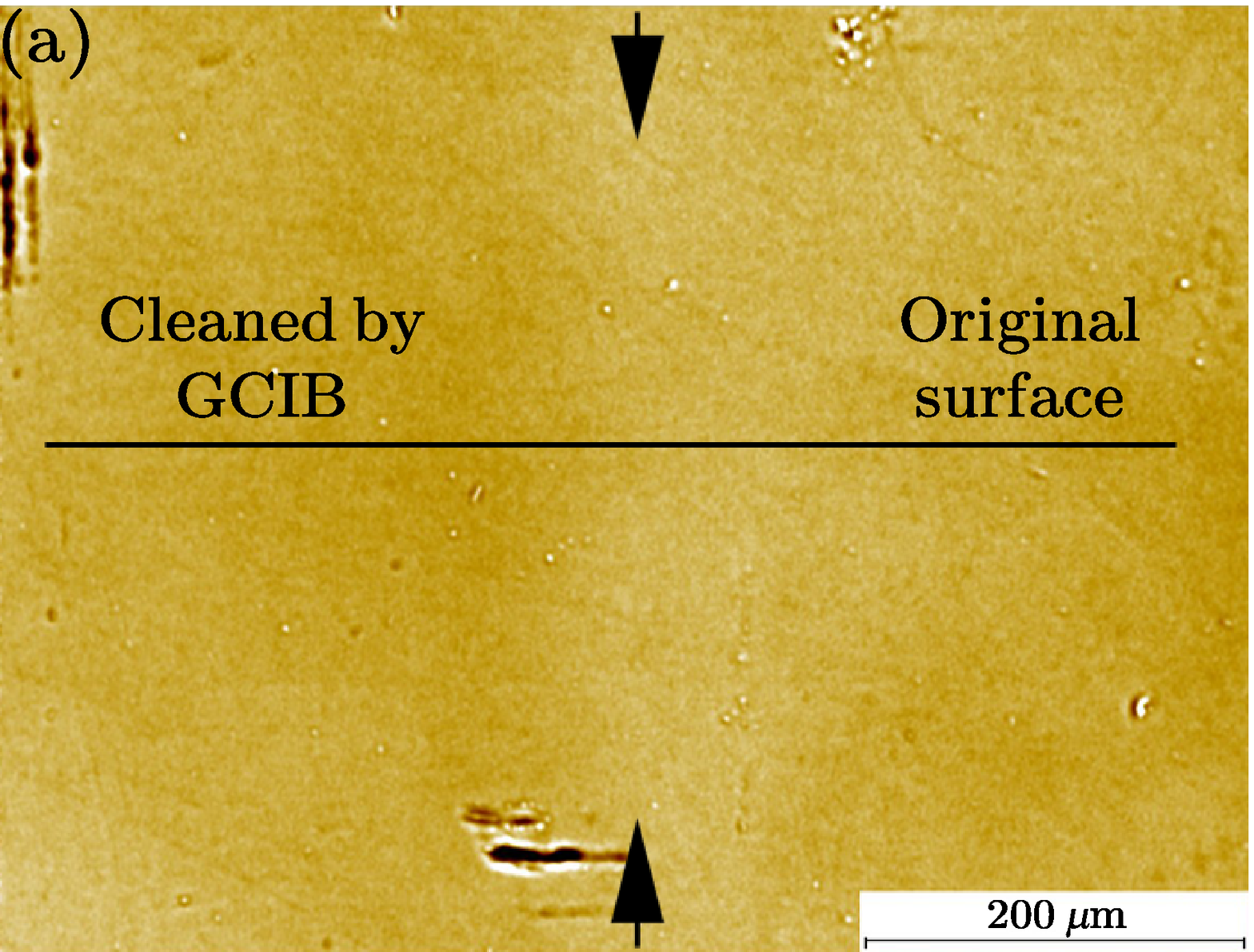}

\

\includegraphics[width=6cm]{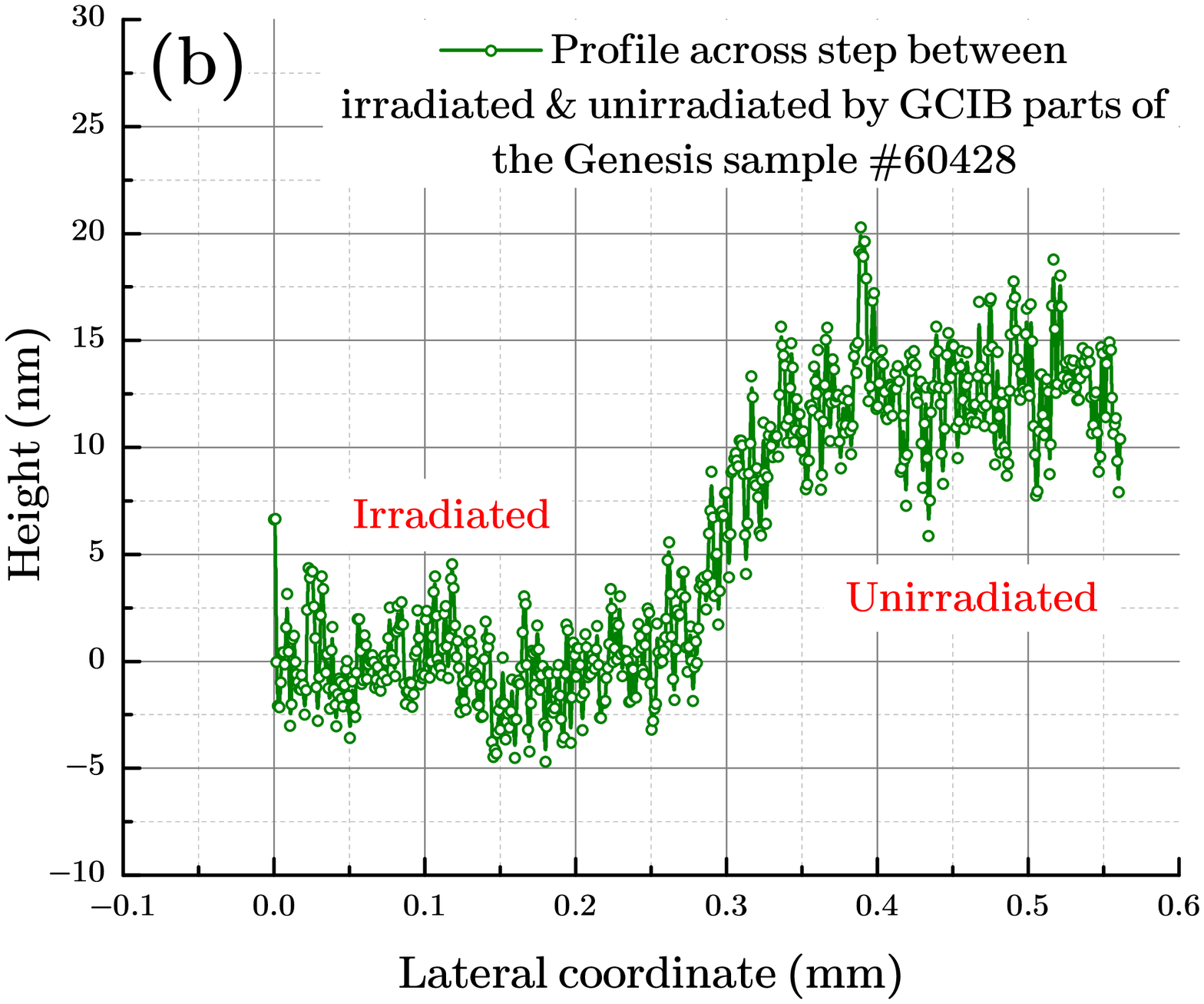}
\caption{(a) Pseudocolor surface of the Genesis Si solar wind
collector coupon 60428 "cleaned" by GCIB. Black line is the
direction along which cross section plotted in (b) of the figure
was measured. Vertical black arrows indicate the separation
between processed/cleaned and original surfaces. (b) WLI cross
sectional profile gives the precise thickness of the removed layer
over the irradiated surface area } \label{fig:5}
\end{figure}

\section{Conclusions}
\label{sec:4}

The benefits of the white light optical profilometry based on a
Mirau interferometer were demonstrated when it is applied to
problems of quantitative characterization of ion sputtered
surfaces. The key advantages of this technique are high depth
resolution in combination with flexible lateral field-of-view and
the capability of true three-dimensional surface topography
reconstruction.

Examples to prove the power of this method were provided here. In
particular, it was demonstrated how to use the WLI approach to
determine sputtering yields of copper and silicon irradiated by
ultralow energy argon ions over confined eroded area of controlled
geometry. Such measurements can be done both on focused static
beam spots ($\sim$10 $\mu$m dia.) and on mm-scale raster scan
craters with high extent of averaging the sputtering
characteristic. In addition, the WLI technique can significantly
help with alignment of ion columns with multiple overlapping or
superimposed ion (or ion and laser) beams, as demonstrated in
presented example with the dual-beam sputter depth profiling.

Thus, the WLI technique facilitates better fundamental
understanding of sputtering processes at ultra-low energies by
helping to accurately determine sputtering yields (and by
addressing problems of preferential sputtering), and by helping
with precise conversion of ion fluence or sputtering time into
depth. Moreover, it lends scientists an ability to precisely
characterize and, ultimately, to control materials' surface
topography formed by ion sputtering under a wide variation of
conditions (eV to tens of keV impact energy or
atomic/cluster/molecular projectile species), which is a great
benefit for ion sputtering based materials synthesis or
characterization.

\section{Acknowledgments}

The authors would like to thank Dr. James Norem (Argonne National
Laboratory, USA) for providing the Cu monocrystals, and Profs.
Isao Yamada and Noriaki Toyoda (University of Hyogo, Japan) for
GCIB processing of the Genesis sample surface. This work was
supported under Contract No. DE-AC02-06CH11357 between UChicago
Argonne, LLC and the U.S. Department of Energy and by NASA through
grants NNH08AH761 and NNH08ZDA001N.

\end{document}